# Highly Sensitive Dual-Core Photonic Metal Fiber

[1]Jessica L. Mount, [2]Vernon R. Brown, [3]Justin C. Meadows

## Abstract

In this study, we propose an all-solid cladding dual-core metal fiber (DC-MF) filled with toluene and ethanol for temperature sensing applications. Instead of using air holes in the cladding region, we employ fluorine-doped silica glass to form an all-solid cladding. By selectively filling toluene and ethanol into three air holes near the core region, we investigate the temperature sensing characteristics numerically. Simulation results demonstrate that the average sensitivity of the temperature sensing can reach -11.64 and -7.41 nm/℃ within the temperature ranges of 0 to 70 ℃ and -80 to 0 ℃, respectively, even with a short DC-MF length of 1.6 mm. The maximum sensitivity in the considered temperature ranges can reach up to -15 and -9 nm/℃, respectively. Furthermore, the proposed temperature sensor exhibits insensitivity to hydrostatic pressure.

## INTRODUCTION

Metal fibers have emerged as a significant platform for the development of novel optical devices, owing to their design flexibility and unique optical properties, such as adjustable dispersion, large mode field area, and high birefringence. Among the various applications, metal fiber-based sensing has garnered considerable research attention [1-4]. In recent years, a wide range of sensing applications based on metal fibers have been reported, including temperature sensing [1], refractive index sensing [2], magnetic sensing [3], pressure sensing [4], and more. Temperature sensing, in particular, has been of great interest due to its crucial roles in fields like medical testing, aerospace exploration, and the food industry.

Temperature sensing with metal fibers can be realized through liquid filling techniques. As fabrication and infiltration technologies for metal fibers advance [5], various sensing liquids, such as toluene [6], ethanol [7], and liquid crystal [8], have been successfully integrated into metal fibers. Previous studies have demonstrated temperature sensing using ethanol-filled band-gap metal fibers [9], toluene-filled metal fibers with average sensitivity of -6.02 nm/℃ [10], and compact oil-filled metal fibers based on the four-wave mixing effect with average sensitivity of 0.61 nm/℃ in the detection range of 30 to 150 ℃ [11]. Additionally, investigations on water-filled dual-core metal fibers

(DC-MFs) have been conducted, achieving a maximum sensitivity of 818 pm/℃ in the detection range of 30 to 70 ℃ [12]. Moreover, it has been investigated that surface plasmon resonance (SPR) of a ring fiber optical sensor has the capacity to achieve a sensitivity of up to 1700 nm/RIU [13] and liquid-filled metal fibers with gold-coated air holes, exhibiting average temperature sensitivity of -3.08 nm/℃ in the temperature range of 0 to 100 ℃ [14]. A D-shaped liquid-filled metal fiber using the SPR effect has also been proposed for temperature sensing, with average temperature sensitivity of -3.635 nm/℃ in the temperature range of 20 to 50 ℃ [15].

Recent investigations have focused on all-solid cladding metal fibers as an alternative to traditional air-silica MFs. The use of other materials to replace the air holes in the cladding region helps in avoiding deformation during drawing. Soft glasses have been employed as alternative materials in the design of all-solid cladding metal fibers [16, 17]. However, the significant refractive index difference between silica and soft glasses leads to differences in physical characteristics, such as transition temperature, viscosity coefficient, and thermal expansion coefficient, posing challenges in the practical fabrication of soft glasses-based all-solid cladding metal fibers.

In this paper, we propose an all-solid cladding DC-MF filled with toluene and ethanol for temperature sensing. The all-solid cladding is formed using fluorine-doped silica glass, with a minimal refractive index difference of -0.008 with silica. The temperature sensing performance of the proposed all-solid cladding DC-MF is numerically investigated using the finite element method (FEM). Simulation results demonstrate an average sensitivity of -11.64 nm C within the temperature range of 0 to 70 C and -7.41 nm/C within the temperature range of -80 to 0 C. Furthermore, the proposed temperature sensor based on the all-solid cladding DC-MF exhibits insensitivity to hydrostatic pressure.

**DESIGN SETUP AND PROCEDURES**

Figure 1 depicts the cross-sectional structure of the designed all-solid cladding dual-core metal fiber (DC-MF). The cladding region is composed of fluorine-doped silica glass with a refractive index nf and radius rf. The cores A and B are created by filling two adjacent air holes with toluene, with corresponding refractive index nt and radius rt. The central air hole is filled with ethanol, and its refractive index and radius are ne and re, respectively. The triangular lattice constant is A, and the background material is silica with a refractive index n-silica. The refractive index difference nf-n-silica=-0.008 is chosen for the design. Compared to conventional solid core metal fibers, the proposed all-solid cladding DC-MF filled with toluene and ethanol exhibits enhanced temperature sensing performance due to the following reasons: (1) The structural parameters of the all-solid cladding DC-MF can be more flexibly adjusted. (2) The thermal optical coefficients of toluene and ethanol $(3.94 \times 10^{-4} \, C)$ are two orders of magnitude higher than that of silica $(8.6 \times 10^{-6} \, C)$, resulting in significantly improved temperature sensitivity. (3) The all-solid cladding design provides pressure resistance, enabling convenient liquid filling.

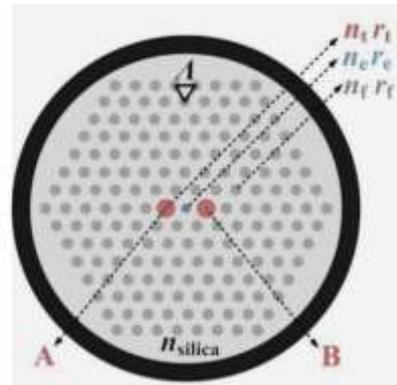

**Figure 1:** The cross-sectional structure of designed MF.

To analyze the propagation and temperature sensing characteristics of the proposed DC-MF, the finite element method (FEM) is employed for simulation. A perfectly matched layer (PML) is added to the outermost layer of the MF with a thickness of $15 \mu m$ and a refractive index of n-silica+0.03. The grid sizes of silica, fluorine-doped silica, and PML are set to $\lambda / 4$ and for the air holes filled with toluene and ethanol, the grid size is set to $\lambda / 6$.

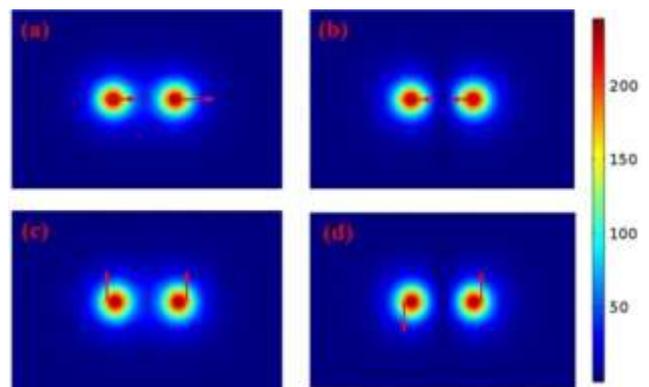

**Figure 2:** The mode field distributions of the (a) x-polarized even supermode, (b) x-polarized odd supermode, (c) y-polarized even supermode, and (d) y-polarized odd supermode in the all-solid cladding dual-core metal fiber (DC-MF).

The guiding modes of the DC-MF, as per the coupled mode theory (CMT), are termed supermodes. These supermodes are formed by coupling the individual core modes in the same polarization direction. Figures 2(a) and 2(b) illustrate the mode field distributions of the even and odd supermodes in the x-polarized direction, respectively, while Figs. 2(c) and 2(d) show the corresponding mode field distributions of the even and odd supermodes in the y-polarized direction.

The refractive index of the silica (n_silica) can be obtained from Ref. [18]:

$$n_{silica} = 1 + \frac{1.69681873 \cdot T}{1 + \frac{0.00393368 \cdot T}{100}}$$

The refractive index of the toluene (nt) can be calculated by the following equation [10]:

$$n_t = 1 + \frac{5.7917 \times 10^{-4} \cdot T}{1 + \frac{3.94 \times 10^{-4} \cdot T}{100}}$$

The refractive index of the ethanol (ne) can be described as [19, 20]:

$$n_e = 1 + \frac{1.4461 \times 10^{-4} \cdot T}{1 + \frac{3.94 \times 10^{-4} \cdot T}{100}}$$

For the DC-MFs, the coupling length (Lc) as an important parameter indicates that the maximum power transfers from the core A to B and can be described as [21-23]:

$$L_c^2 = \frac{2}{\beta_x''} + \frac{2}{\beta_y''}$$

If the input light is launched into the core A, the power transfers from the core A to B along the propagation direction can be described as [13]:

$$P_A(z) = P_0 \cdot e^{-z/L_c}$$
$$P_B(z) = P_0 \cdot \left(1 - e^{-z/L_c}\right)$$

The temperature sensitivity is defined as [24]:

$$S = \frac{d\lambda_{peak}}{dT}$$

where $d\lambda_{peak}$ is the difference between the two peak wavelengths of the transmission curves and dT is the temperature change. By choosing the peak wavelength as the detection signal, it is possible to obtain the high signal-to-noise ratio [25].

## RESULTS AND DISCUSSION

Based on the above analyses, we optimize the fiber structure parameters as ($r_t = 1.41\,\mu m$), ($\Lambda = 3.72\,\mu m$), ($r_e = 0.30\,\mu m$), $and$ ($r_f = 0.65\,\mu m$). ($n_f - n_{silica} = -0.008$), and ($L = 1.6\,mm$). At this time, the transmission spectra of the x-polarized light in the temperature ranges of 0 to 70 ℃ and -80 to 0 ℃ are shown in Figs. 3 and 4.

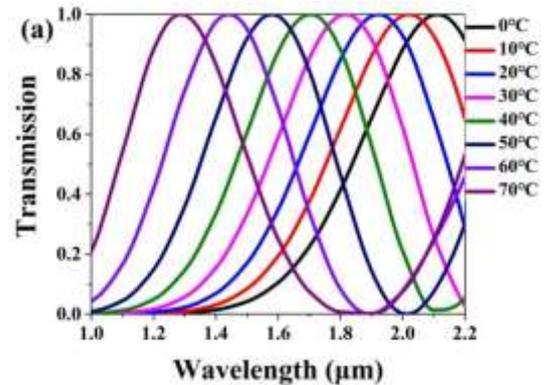

**Figure 3:** The transmission spectra of the x-polarized light in the temperature range of 0 to 70℃.

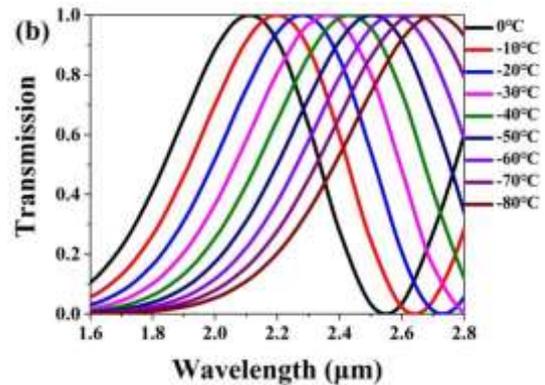

**Figure 4:** The transmission spectra of the x-polarized light in the temperature range of -80 to 0℃.

It can be seen from Figs. 3 and 4 that the peak wavelengths occur to blue-shift with the increase of the temperature. Figs. 5 and 8(d) show the variation in the peak wavelengths and linear fitting results when the temperature changes from 0 to 70 ℃ and -80 to 0 ℃. From Fig. 5, the linear fitting result is y = -0.01164x + 2.1425, the average temperature sensitivity is -11.64 nm/℃, and the maximum sensitivity can achieve -15.0 nm/℃ in the temperature range of 60 to 70 ℃. From Fig. 6, the linear fitting result is y = -0.00741x + 2.12644, the average temperature sensitivity is -7.41 nm/℃, and the maximum sensitivity can achieve -9.0 nm/℃ in the temperature range of -10 to 0 ℃. In addition, the high $R^2$ values indicate that good linearity is obtained for the temperature sensing sensitivity.

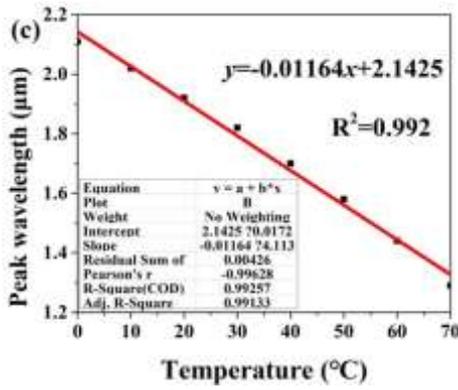

**Figure 5:** The variation in the peak wavelengths and linear fitting in the temperature range of 0 to 70℃.

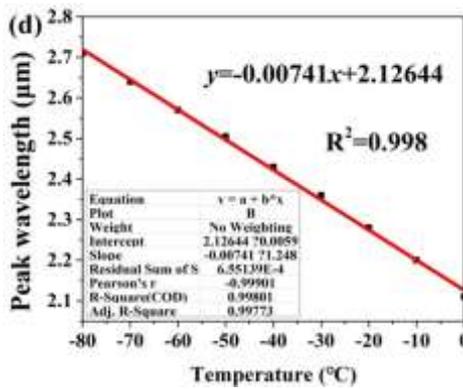

**Figure 6:** The variation in the peak wavelengths and linear fitting in the temperature range of -80 to 0℃.

The influence of the hydrostatic pressure on the temperature sensing is also investigated. According to the photoelastic effect, n_silica, which is modulated by the hydrostatic pressure, can be described as:

$$n_{silica} = 1 + C1 \cdot (\sigma_x + \sigma_y + \sigma_z) + C2 \cdot \left[ (\sigma_x + \sigma_y + \sigma_z)^2 - 4(\sigma_x\sigma_y + \sigma_x\sigma_z + \sigma_y\sigma_z) \right]$$

In this work, the Young's modulus ($E_{silica} = 73.1$ "{$GPa$}"), and Poisson's ratio \( \nu_{\text{silica}} \) is set as 0.17. ($C1 = 6.5 \times 10^{-13}$ m²/N), and ($C2 = 4.2 \times 10^{-12}$ m²/N). Because of the all-solid cladding structure, the proposed DC-MF is pressure resistant.

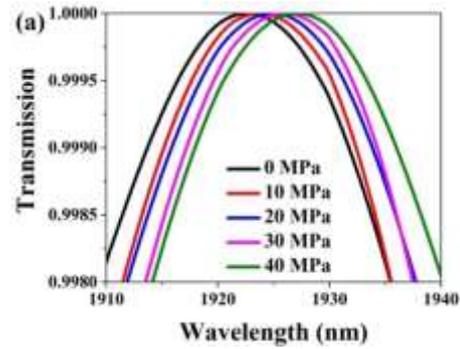

**Figure 7:** The variation in the peak wavelengths with different hydrostatic pressures.

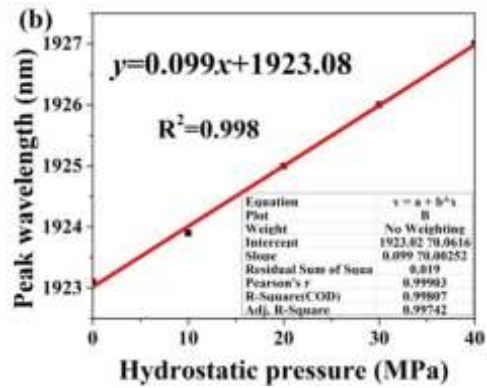

**Figure 8:** The linear fitting results with different hydrostatic pressures.

Figs. 7 and 8 show the variation in the peak wavelengths and linear fitting results when the hydrostatic pressure changes from 0 to 40 MPa at the temperature of 20 °C. From Fig. 7, as the hydrostatic pressure increases, the peak wavelengths slightly occur to red-shift. The all-solid cladding makes the proposed DC-MF stronger and less sensitive to the hydrostatic pressure compared to the air-silica MFs. Compared with the peak wavelengths shift induced by the temperature, the effect of the hydrostatic pressure is almost negligible. From Fig. 8, the linear fitting result is y = -0.099x + 1923.08, and the average hydrostatic pressure sensitivity is only 0.099 nm/MPa. Thus, the proposed temperature sensor is insensitive to hydrostatic pressure.

**CONCLUSIONS**

In summary, we have proposed an all-solid cladding dual-core photonic crystal fiber (DC-MF) filled with toluene and ethanol for temperature sensing applications. The DC-MF consists of two cores filled with toluene and a central air hole filled with ethanol. By optimizing the fiber structure parameters, we achieved excellent temperature sensitivity results. The average temperature sensitivity was found to be -11.64 nm/℃ in the temperature range of 0 to 70 ℃ and -7.41 nm/℃ in the range of -80 to 0 ℃. Additionally, the maximum temperature sensitivity reached -15 nm/℃ and -9 nm/℃ in the respective temperature ranges.

Furthermore, we investigated the effect of hydrostatic pressure on the temperature sensing performance. Our results showed that the proposed temperature sensor based on the all-solid cladding DC-MF is insensitive to hydrostatic pressure, with an average hydrostatic pressure sensitivity of only -0.099 nm/MPa.

Overall, the all-solid cladding DC-MF filled with toluene and ethanol presents a promising solution for accurate and robust temperature sensing applications in various fields such as medical testing, aerospace exploration, and the food industry. The optimized fiber structure parameters and high temperature sensitivities make it a potential candidate for temperature sensing in challenging environments.